\documentclass[conference, 9pt]{IEEEtran}

\usepackage{graphicx,subfigure}
\usepackage{amsmath,amsfonts,amssymb}
\usepackage{multirow}

\newtheorem{defn}{Definition}
\newtheorem{thm}{Theorem}
\newtheorem{lem}{Lemma}

\begin{document}

\title{On Unconditionally Secure Multiparty Computation for Realizing Correlated Equilibria in Games}

\author{\IEEEauthorblockN{Ye Wang, Shantanu Rane}
\IEEEauthorblockA{\emph{Mitsubishi Electric Research Laboratories}\\
Cambridge, MA 02139, USA.\\
\{yewang,rane\}@merl.com\\}
\and
\IEEEauthorblockN{Prakash Ishwar}
\IEEEauthorblockA{\emph{Boston University}\\
Boston, MA 02215, USA.\\
pi@bu.edu}}

\maketitle

\begin{abstract}
In game theory, a trusted mediator acting on behalf of the players can enable the attainment of correlated equilibria, which may provide better payoffs than those available from the Nash equilibria alone. We explore the approach of replacing the trusted mediator with an unconditionally secure sampling protocol that jointly generates the players' actions. We characterize the joint distributions that can be securely sampled by malicious players via protocols using error-free communication. This class of distributions depends on whether players may speak simultaneously (``cheap talk'') or must speak in turn (``polite talk''). In applying sampling protocols toward attaining correlated equilibria with rational players, we observe that security against malicious parties may be much stronger than necessary. We propose the concept of secure sampling by rational players, and show that many more distributions are feasible given certain utility functions. However, the payoffs attainable via secure sampling by malicious players are a dominant subset of the rationally attainable payoffs.
\end{abstract}

\begin{IEEEkeywords} game theory, secure sampling, correlated randomness \end{IEEEkeywords}

\section{Introduction}

The fields of game theory and secure computation are both primarily concerned with the interactions of mutually untrusting parties.
Game theory studies the behavior of \emph{rational} players that seek to maximize
their personal payoffs as derived from the outcome of their interaction~\cite{osborne94book}.
Equilibria are an important concept in understanding how rational players may attain and enforce desirable payoffs.
Secure computation concerns the design of interactive protocols that allow the parties to perform computations while retaining privacy of their inputs~\cite{yao82focs,micali92crypto}.
These parties may be \emph{semi-honest}, i.e., they follow the protocol but may attempt to infer information about other parties, or they may be \emph{malicious}, i.e., they may 
arbitrarily deviate from the protocol in an attempt to subvert its privacy and/or correctness.
Loosely speaking, both equilibria and secure protocols are means to ensure resilience against, respectively, selfish rationality and arbitrary maliciousness.
Thus, there are interesting philosophical and theoretical connections that have motivated cross-pollination of ideas across these disciplines~\cite{dodis07bookchapter,katz08toc}.
This work explores the application of secure computation in attaining equilibria with rational players.

Informally, consider players that observe some randomness (e.g., a coin flip) that assigns correlated actions to each player.
Such a strategy is a correlated equilibrium if no player has an incentive to unilaterally deviate from its assigned action~\cite{osborne94book}.
The expected payoffs of this correlated equilibrium may be better than those obtained by any Nash equilibrium. 
One method for realizing a correlated equilibrium is to use a trusted mediator who generates the correlated actions for the players.
However, we are concerned with situations in which a trusted mediator is not available.
In this case, secure computation suggests a solution: The players can run a secure protocol that simulates the mediator while guaranteeing the correctness of correlated action distribution, and the privacy of the actions assigned to each player.
This specific secure computation problem of securely generating samples from a joint distribution is known as secure sampling.
While much of the existing work has considered settings with computationally bounded players, we study the feasibility of simulating the mediator with unconditional (information-theoretic) security guarantees.

The remainder of this paper is organized as follows: In Sec.~\ref{sec:prelims}, we briefly review background material on correlated equilibria and secure sampling with semi-honest parties.
Sec.~\ref{sec:mainresults} contains our main result, which characterizes the class of joint distributions that can be securely sampled by malicious players, and discusses the payoffs attainable via the secure sampling of correlated equilibria.

\section{Preliminaries}
\label{sec:prelims}

\subsection{Games and Equilibria}
\label{sec:gametheory}

For simplicity of exposition, our development will focus on two-player strategic games with complete information and finite action sets\footnote{Strictly speaking, our analysis inherently must consider extended games with infinite action sets in order to analyze mixed strategies, however, the core game motivating the development is assumed to have finite action sets.}.
Such games are given by a pair of finite action sets $\mathcal{X}, \mathcal{Y}$, and utility functions $u_1, u_2: \mathcal{X} \times \mathcal{Y} \rightarrow \mathbb{R}$.
In an execution of the game, the first player (who we will call Alice) plays an action $X \in \mathcal{X}$, and the second player (who we will call Bob) plays an action $Y \in \mathcal{Y}$.
The actions $(X,Y)$, together called an {\em action profile}, may be randomly chosen, but are revealed simultaneously.
Each realization $(x,y) \in \mathcal{X} \times \mathcal{Y}$ constitutes an outcome in the game, with the utility functions $u_1(x,y)$ and $u_2(x,y)$ quantifying the respective payoffs for Alice and Bob under that outcome.

The objective of a {\em rational} player is to play in a manner that maximizes his or her personal utility.
The concept of equilibria is important toward understanding how rational players may behave in playing such games~\cite{osborne94book}.
We will use the definition of {\em mixed-strategy} Nash equilibrium and subsequently refer to them as Nash equilibria\footnote{The {\em pure-strategy} Nash equilibria are the subset of mixed-strategy Nash equilibria where the action profiles are deterministic.}.

\begin{defn} \label{def:nasheq}
A pair of distributions $(P_X, P_Y)$ on $\mathcal{X}$ and $\mathcal{Y}$, respectively, is a {\em mixed-strategy Nash equilibrium} of game $(\mathcal{X}, \mathcal{Y}, u_1, u_2)$ if, and only if, for all distributions $P_{X'}$ on $\mathcal{X}$ and $P_{Y'}$ on $\mathcal{Y}$, we have that $E[u_1(X,Y)] \geq E[u_1(X',Y)]$ and $E[u_2(X,Y)] \geq E[u_2(X,Y')]$,
where the random variables $(X,X',Y,Y') \sim P_X P_{X'} P_Y P_{Y'}$.
\end{defn}

An interpretation of Nash equilibria is that they are strategy profiles to which, if the players have committed, there is no incentive for either player to unilaterally deviate in their action.
For the finite action set games that we consider, the set of mixed-strategy Nash equilibria is non-empty~\cite{osborne94book}.

\begin{defn} \label{def:correq}
A joint distribution $P_{X,Y}$ on $\mathcal{X} \times \mathcal{Y}$ is a {\em correlated equilibrium} of game $(\mathcal{X}, \mathcal{Y}, u_1, u_2)$ if, and only if, for all random variables $X'$ and $Y'$ such that $X' - X - Y$ and $Y' - Y - X$ form Markov chains with $(X,Y) \sim P_{X,Y}$, we have that $E[u_1(X,Y)] \geq E[u_1(X',Y)]$ and $E[u_2(X,Y)] \geq E[u_2(X,Y')]$.
\end{defn}

The Nash equilibria are also correlated equilibria (with $P_{X,Y} = P_X P_Y$), and hence the set of correlated equilibria is also non-empty.
An interpretation for correlated equilibria is that if a trusted mediator chose actions $(X,Y) \sim P_{X,Y}$ and revealed each action only to its respective player to play, then there would be no incentive for either player to unilaterally deviate from their given action\footnote{Formally speaking, a Nash equilibrium of the extended game, where nature (acting as the mediator) first chooses and distributes actions to the players before they play, would be both players following their given action.}.
For the correlated equilibria that are also Nash equilibria, the mediator is not necessary, since players could choose their actions independently.
Thus, of particular interest are the correlated equilibria that are not also Nash equilibria, hence seeming to require a mediator to be realized.

For a given game, we say that a pair $(p_1,p_2) \in \mathbb{R}^2$ is a {\em Nash payoff} (or {\em correlated payoff}) if, and only if, the game has a Nash (respectively, correlated) equilibrium with expected payoffs $(p_1,p_2)$.
The sets of Nash and correlated equilibria correspond to sets of Nash and correlated payoffs.
 
\begin{figure}
\centering
\subfigure[``Battle-of-the-Sexes'']{
  \includegraphics[width=1.1in]{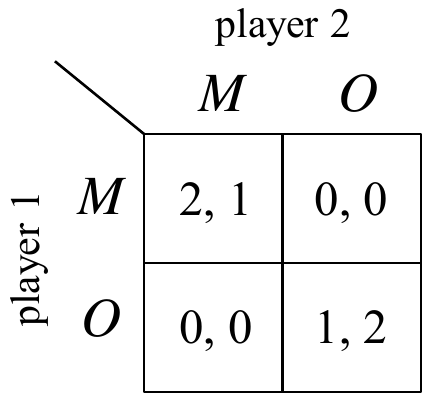}
  \label{fig:bos}}
\quad \quad
\subfigure[``Chicken-or-Dare'']{
  \includegraphics[width=1.1in]{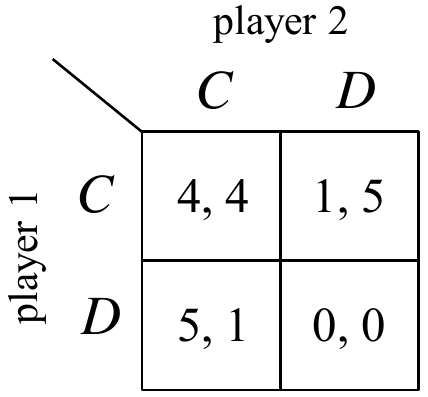}
  \label{fig:cod}}
\caption{Common examples of two-player games}
\label{fig:games}
\end{figure}

Some examples of well-known games are given in Fig.~\ref{fig:games}.
The ``Battle-of-the-Sexes'' game, given in Fig.~\ref{fig:bos}, has three Nash equilibria ($P_X(M)=P_Y(M)=1$, $P_X(O)=P_Y(O)=1$, and $P_X(M)=P_Y(O)=2/3$) and a correlated equilibrium for every $\lambda \in [0,1]$ ($P_{X,Y}(x,y) = \lambda \mathbf{1}(x=y=M) + (1 - \lambda) \mathbf{1}(x=y=O)$).
The ``Chicken-or-Dare'' game, given in Fig.~\ref{fig:cod}, has three Nash equilibria ($P_X(C)=P_Y(D)=1$, $P_X(D)=P_Y(C)=1$, and $P_X(D)=P_Y(D)=1/2$) and a correlated equilibrium ($P_{X,Y}(x,y) = 1/3$ for $(x,y) \in \{(C,C), (C,D), (D,C)\}$).

\subsection{Secure Sampling Protocols}
\label{sec:securesampling}

In the problem of secure two-party sampling, the objective is to design a {\em secure} interactive protocol that allows the two parties to generate correlated randomness according to a given joint distribution, where security means that {\em correctness} (in terms of the randomness actually matching the desired distribution) and {\em privacy} (in terms of revealing nothing about either output except what is inherent due to correlation) of the outputs are assured.

A two-party {\em interactive protocol} is a pair of algorithms designed to interact over several rounds via communication channels. 
In this work, we will restrict our scope to protocols using only error-free channels.
We will further consider two specific cases of error-free communication. Using the game-theoretic terminology, they are {\em cheap talk}, where parties may exchange messages simultaneously, and {\em polite talk}, where parties must take turns in exchanging messages\footnote{In each round of cheap talk, neither party is able to see the other's message before sending their own. In either scenario, we assume some weak synchronization (i.e., a ``timeout'' mechanism) preventing a missing message from holding up the protocol.}.
In between the rounds of interaction, the algorithms may generate independent randomness, perform local computations, and compute the next message to be sent from everything observed in previous rounds.
After the interaction has terminated, each party produces an output that may be computed from everything that they have observed during the protocol.
Unlike previous work (see~\cite{WangIshwarISIT11}), we will not place a finite bound on the rounds of interaction and instead allow protocols that potentially last a random, unbounded number of rounds, provided that they terminate almost surely when at least one of the parties is honest.

Motivated by the application to rational players, we are primarily interested in the security of protocols against malicious parties that may arbitrarily deviate from the protocol in an attempt to undermine correctness and privacy.
The following definition for secure sampling stipulates that the correctness and privacy of the honest parties be maintained if there is at least one honest party.
This definition is an adaptation of a definition for secure {\em computation} in~\cite{CrepeauSSW-Eurocrypt06-ITSecCond2PSFE}, where it is shown to be equivalent to simulating a trusted mediator.

\begin{defn} \label{def:secactive} (see~\cite{CrepeauSSW-Eurocrypt06-ITSecCond2PSFE})
A two-party protocol for sampling $P_{X,Y}$ is secure against malicious parties if, and only if, for any execution generating outputs $(U,V)$, we have that:
\begin{enumerate}
\item When both parties are honest, $(U,V) \sim P_{X,Y}$.
\item \label{itm:AliceBad} When Bob is honest, there exists random variable $\overline{X}$ such that
$(\overline{X}, V) \sim P_{X,Y}$, and
$U - \overline{X} - V$ forms a Markov chain.
\item \label{itm:BobBad} When Alice is honest, there exists random variable $\overline{Y}$ such that
$(U ,\overline{Y}) \sim P_{X,Y}$, and
$V - \overline{Y} - U$ forms a Markov chain.
\end{enumerate}
\end{defn}

Note that in conditions~\ref{itm:AliceBad} and~\ref{itm:BobBad} of Defn.~\ref{def:secactive}, one player is fixed to be honest while the other can arbitrarily deviate from the protocol. This may include the deviating party changing their output to be any information gathered about the other party's output, and hence the Markov chain requirements capture privacy in the sense that a deviating party can only extract information that would have been inherent to a hypothetical output (i.e., $\overline{X}$ or $\overline{Y}$).

The next definition provides the conditions for security against {\em semi-honest parties} where the parties execute the protocol honestly but may attempt to infer additional information from what they observe.
This is a weaker security condition implied by that for malicious parties (see Lem.~\ref{lem:ActiveSecStronger}) and will be useful for proving impossibility results for the malicious party case.
The conditions require that, for an honest execution of the protocol, the outputs are correctly generated and that the {\em view} of each player (consisting of all  intermediate computations, local randomness, and messages exchanged) does not reveal any information not already derivable from each player's own output.

\begin{defn} \label{def:secpassive} (see~\cite{WangIshwarISIT11})
A two-party protocol for sampling $P_{X,Y}$ is secure against semi-honest parties if, and only if, for any honest execution generating outputs $(U,V)$, we have that:
\begin{enumerate}
\item $(U,V) \sim P_{X,Y}$.
\item $\mathrm{View}_A - U - V$ forms a Markov chain, where $\mathrm{View}_A$ denotes the view of Alice.
\item $\mathrm{View}_B - V - U$ forms a Markov chain, where $\mathrm{View}_B$ denotes the view of Bob.
\end{enumerate}
\end{defn}

\subsection{Background Results}

The feasibility of secure two-party sampling with semi-honest parties has been previously characterized in~\cite{WangIshwarISIT11}.

\begin{thm} \label{thm:PassiveFeasibility} (see~\cite[Cor.~1]{WangIshwarISIT11})
Given polite and/or cheap talk channels, there exist two-party protocols\footnote{This result assumed bounded rounds of communication, however the converse can be extended to unbounded communication. The corresponding achievability result trivially extends to the unbounded case.} for securely sampling $P_{X,Y}$ against semi-honest parties if, and only if,
\begin{equation} \label{eqn:CommInfo}
I(X;Y) = C(X;Y) := \min_{W: I(X;Y|W) = 0} I(X,Y ; W),
\end{equation}
where $C(X;Y)$ is called the Wyner common information~\cite{Wyner-75-CommonInfo}.
\end{thm}

In general, $I(X;Y) \leq C(X;Y)$, with equality holding only for the special class of random variables $(X,Y)$ where a minimizing $W$ in~(\ref{eqn:CommInfo}) is the {\em ergodic decomposition} of $(X,Y)$~\cite{GacsKorn-73-CommonInfo, AhlsKorn-74-CommonInfo}.
The ergodic decomposition is given by first uniquely labeling the connected components in the graphical representation\footnote{The graphical representation of $P_{X,Y}$ is the bipartite graph with an edge between $(x,y) \in \mathcal{X} \times \mathcal{Y}$ iff $P_{X,Y}(x,y) > 0$.} of $P_{X,Y}$, and then assigning $W$ to the label of the connected component in which $(X,Y)$ falls.
Clearly, the ergodic decomposition is also a deterministic function of either $X$ or $Y$ alone.
As a consequence of the above, the following lemma provides a condition equivalent to $I(X;Y) = C(X;Y)$, in terms of a simple property of the ergodic decomposition of $(X,Y)$.

\begin{lem}
Given $P_{X,Y}$, $I(X;Y) = C(X;Y)$ if, and only if, for $W$, the ergodic decomposition of $(X,Y)$, the Markov chain $X - W - Y$ holds.
\end{lem}

A widely studied special case of the secure sampling with malicious parties is that of generating an unbiased coin-flip, that is, $P_{X,Y}(x,y) = 0.5 \cdot \mathbf{1}(x=y)$ with $\mathcal{X} = \mathcal{Y} = \{0,1\}$.
It is well known that it is impossible to perform a secure coin flip given only {\em bounded} polite talk~\cite{Cleve-STOC86-CoinFlips, HanggiW-TCC11-CoinTossBounds}.
A tight characterization of the tradeoff between protocol reliability (in an honest execution) and the potential bias that can be introduced by a cheating party is given in~\cite{HanggiW-TCC11-CoinTossBounds}, which shows that at least one party must always be able to significantly bias the coin flip.
On the other hand, given cheap talk, performing a secure coin flip is trivial; each player chooses a uniform bit independently, the players simultaneously exchange bits via cheap talk, and assign the coin flip as the binary XOR of the bits\footnote{If one party chooses not to send a valid bit, the other will simply set the coin flip to their own bit.}.
Also, given cheap talk, this procedure can be extended to securely sample a series of independent coin flips or any general discrete random variable.

\section{Main Results}
\label{sec:mainresults}

Our main result is the characterization of the set of distributions that can be securely sampled by malicious parties via protocols using either cheap talk or only polite talk.
The feasibility region boils down to ``separable'' distributions (where $I(X;Y) = C(X;Y)$) for the cheap talk case, and ``trivial'' distributions (where $X$ and $Y$ are independent) for the polite talk case.
For a given game, the correlated equilibria (and their corresponding payoffs) that are in this feasible set can be realized via a secure protocol replacing a trusted mediator.
However, we ask whether security against malicious parties is too strong a requirement, and discuss the correlated equilibria and payoffs attainable by rational players.

\subsection{Secure Protocols for Sampling}

\begin{lem} \label{lem:ActiveSecStronger}
If a two-party protocol for sampling $P_{X,Y}$ is secure against malicious parties (see Defn.~\ref{def:secactive}), then it is also secure against semi-honest parties (see Defn.~\ref{def:secpassive}).
\end{lem}

\IEEEproof
Consider an ``almost honest'' execution of the protocol, where the parties follow the protocol honestly, except that they append their full views to their honest outputs.
Let these outputs be denoted by $U' := (\mathrm{View}_A, U)$ and $V' := (\mathrm{View}_B, V)$ respectively, where $(U, V)$ are the outputs and $(\mathrm{View}_A, \mathrm{View}_B)$ are the views produced by an honest execution.
The security of the protocol against malicious parties implies that $(U,V) \sim P_{X,Y}$, and that there exist $\overline{X}$ and $\overline{Y}$ such that $(\overline{X}, V) \sim P_{X,Y}$, $(U, \overline{Y}) \sim P_{X,Y}$, and the Markov chains $(\mathrm{View}_A, U) - \overline{X} - V$ and $(\mathrm{View}_B, V) - \overline{Y} - U$ hold.
The Markov chain $(\mathrm{View}_A, U) - \overline{X} - V$ implies that
\begin{align*}
0 &= I(\mathrm{View}_A, U; V | \overline{X}) \\
  &= H(V | \overline{X}) - H(V | \overline{X}, \mathrm{View}_A, U) \\
  &\geq H(V | U) - H(V | \mathrm{View}_A, U) = I(\mathrm{View}_A; V | U),
\end{align*}
and hence the Markov chain $\mathrm{View}_A - U - V$ holds. Similarly, it can be shown that the Markov chain $\mathrm{View}_B - V - U$ holds. Therefore, the protocol is secure against semi-honest parties.
\endproof

\begin{thm} \label{thm:ActiveFeasibility}
Given cheap talk channels, there exist two-party protocols for securely sampling $P_{X,Y}$ against malicious players if, and only if, $I(X;Y) = C(X;Y)$.
Secondly, given only polite talk channels, there exist two-party protocols for securely sampling $P_{X,Y}$ against malicious players if, and only if, $P_{XY} = P_X P_Y$, that is, the random variables are independent.
\end{thm}

\IEEEproof {\em (sketch)}
First, we consider the situation with cheap talk.
The converse (``only if'') direction is due to the impossibility of securely sampling $P_{X,Y}$ if $I(X;Y) \neq C(X;Y)$ even against semi-honest parties (see Thm.~\ref{thm:PassiveFeasibility}) which also applies to the malicious case by Lem.~\ref{lem:ActiveSecStronger}.
To show the achievability (``if'') direction, we argue that given $I(X;Y) = C(X;Y)$, a secure protocol can be constructed that first securely generates the ergodic decomposition $W$ of $(X,Y)$ using cheap talk, followed by each party independently generating $X$ and $Y$, respectively, from $W$.
Since $W$ is a function of either $X$ or $Y$ alone, it does not reveal to each party any additional information about the other party's output than their own output.

Next, we consider the situation with only polite talk.
In the converse (``only if'') direction, the semi-honest converse requiring $I(X;Y) = C(X;Y)$ similarly applies as above. Thus, given a protocol securely sampling $P_{X,Y}$, we must have that the ergodic decomposition $W$ of $(X,Y)$ satisfies the Markov chain $X - W - Y$. We argue by contradiction that $W$ must be deterministic, and hence that $P_{X,Y} = P_X P_Y$ following from the Markov chain.
If $W$ is not deterministic, then there exists a partitioning of its alphabet $\mathcal{W}$ into $\mathcal{W}_0$ and $\mathcal{W}_1$ such that $\Pr(W \in \mathcal{W}_0), \Pr(W \in \mathcal{W}_1) \in (0,1)$.
Thus, since $W$ is a function of either $X$ or $Y$, one can convert the secure protocol for $P_{X,Y}$ into a protocol for a secure biased coin flip\footnote{This coin flip would be inherently biased, but secure in the sense that no player can alter that bias.} by mapping $X$ (or $Y$) to the $Z \in \{0,1\}$ where $W \in \mathcal{W}_Z$.
It follows from~\cite{HanggiW-TCC11-CoinTossBounds} that generating such a secure coin flip from polite talk is impossible\footnote{The proof of~\cite{HanggiW-TCC11-CoinTossBounds} assumes protocols with finitely bounded interaction, however, this result can be extended, albeit with a non-trivial argument that we must omit due to space, to protocols with unbounded interaction but almost sure termination given at least one honest party.}, and hence $W$ must be deterministic.
The achievability (``if'') direction is immediate for $P_{X,Y} = P_X P_Y$ via the trivial protocol with no interaction and the parties independently generating $X$ and $Y$.
\endproof

The secure sampling feasibility results of Thm.~\ref{thm:PassiveFeasibility} and Thm.~\ref{thm:ActiveFeasibility} are summarized in Table~\ref{tab:secsampresults}.

\begin{table}
\center
\renewcommand{\arraystretch}{1.5}
\begin{tabular} {| r || c | c |}
\hline
 & {\bf Semi-honest Parties} & {\bf Malicious Parties} \\
\hline \hline
{\bf Polite Talk} & \multirow{2}{*}{$I(X;Y) = C(X;Y)$} & $P_{X,Y} = P_X P_Y$ \\
\cline{1-1}\cline{3-3}
{\bf Cheap Talk} &  & $I(X;Y) = C(X;Y)$ \\
\hline
\end{tabular}
\caption{Feasibility Conditions for Secure Two-Party Sampling}
\label{tab:secsampresults}
\end{table}

\subsection{Rational Protocols for Games}

Secure sampling protocols can be applied toward realizing correlated equilibria when players lack a trusted mediator but are able to first communicate via cheap talk or polite talk.
The following definition provides sufficient conditions for a protocol that would allow \emph{rational} players to realize a given correlated equilibrium in lieu of a mediator\footnote{Formally speaking, for any correlated equilibrium for which a rationally secure protocol exists, a Nash equilibrium of the extended game, where parties may first interact via cheap or polite talk and then play moves, would be both players honestly executing that protocol and playing the outputs generated.}.

\begin{defn} \label{def:secrational}
Let $u_1, u_2: \mathcal{X} \times \mathcal{Y} \rightarrow \mathbb{R}$ be the payoffs in a two-player game with a correlated equilibrium $P_{X,Y}$.
A two-party protocol for sampling $P_{X,Y}$ is secure against rational players if, and only if, for any execution generating outputs $(U,V)$ with support in $\mathcal{X} \times \mathcal{Y}$\footnote{Without loss of generality, we need only consider deviations that still generate outputs in the appropriate alphabets, since the ultimate choice of action can be subsumed into the deviation from the protocol.}, we have that:
\begin{enumerate}
\item When both parties are honest, $(U,V) \sim P_{X,Y}$.
\item When Bob is honest, $E[u_1(U,V)] \leq E[u_1(X,Y)]$, where $(X,Y) \sim P_{X,Y}$.
\item When Alice is honest, $E[u_2(U,V)] \leq E[u_2(X,Y)]$, where $(X,Y) \sim P_{X,Y}$.
\end{enumerate}
\end{defn}

Protocols that are secure against malicious parties ensure that any deviation cannot subvert the correctness or privacy of the sampling.
Hence, such protocols would be sufficient for realizing correlated equilibria for rational players (see Lem.~\ref{lem:MalSecImplRatSec} below).

\begin{lem} \label{lem:MalSecImplRatSec}
Given a two-player game with a correlated equilibrium, $P_{X,Y}$, if a two-party protocol for sampling $P_{X,Y}$ is secure against malicious parties (see Defn.~\ref{def:secactive}), then it is also secure against rational players (see Defn.~\ref{def:secrational}).
\end{lem}

\IEEEproof
The first condition of Defn.~\ref{def:secrational} is immediate from Defn.~\ref{def:secactive}.
For the second condition of Defn.~\ref{def:secrational}, Defn.~\ref{def:secactive} requires that for any Alice (including those that only output $U \in \mathcal{X}$), there exists $\overline{X}$ such that $U-\overline{X}-V$ and $(\overline{X},V) \sim P_{X,Y}$.
Since, $P_{X,Y}$ is a correlated equilibrium, we have that $E[u_1(U,V)] \leq E[u_1(\overline{X},V)] = E[u_1(X,Y)]$.
The third condition of Defn.~\ref{def:secrational} follows similarly.
\endproof

However, protocols that are secure for malicious parties may be unnecessarily strong for rational players that will only deviate if it serves their best interests.
Further, as follows from Thm.~\ref{thm:ActiveFeasibility}, only a limited correlated equilibria can be securely sampled by malicious parties.
In particular, given only polite talk, this is limited to only the trivial Nash equilibria.
Hence, a valid question is whether requiring security against only rational players would allow for a larger set of attainable correlated equilibria or corresponding payoffs.

Consider the (somewhat pathological\footnote{Following similar principles, one can also construct less pathological examples exhibiting this significant gap.}) scenario where the payoff functions are constant, and hence all joint distributions over the action profiles are correlated equilibria.
All distributions can be securely sampled against rational players given constant payoffs, using the simple protocol where the first party generates both $(X,Y) \sim P_{X,Y}$, gives $Y$ to the second party, and then each party outputs its respective variable.
However, notice that, for this example, all of the correlated and Nash payoffs are the same, thus the expanded range of equilibria that can be sampled with rational parties does not correspond to an expansion of the attainable correlated payoffs.
Thus, the pertinent comparison appears to be the payoffs attainable by rational players versus those attainable via secure sampling by malicious parties.

The sets of correlated payoffs achievable by rational players using cheap or polite talk have been characterized in~\cite{AumannHart-03-LongCheapTalk}.
Their results, specialized to our scenario, give that the set of the achievable payoffs for rational players with cheap talk is the convex hull of the Nash payoffs, while the set of achievable payoffs with polite talk is the biconvex-span of the Nash payoffs\footnote{The biconvex-span is a subset of the convex hull and consists of the expectations of possible bounded bimartingales that converge almost surely to a Nash payoff (see~\cite{AumannHart-03-LongCheapTalk} for more details).}.
Any payoff in the convex hull of Nash payoffs corresponds to a correlated equilibrium that is a convex combination of the Nash equilibria, i.e., there exist $Z$ such that $P_{X,Y} = \sum_{z \in \mathcal{Z}} P_Z(z) P_{X,Y|Z=z}$, where $P_{X,Y|Z=z} = P_{X|Z=z}P_{Y|Z=z}$ are the Nash equilibria.
The distribution $P_{A,B}$, given by $A := (X,Z)$ and $B := (Y,Z)$, is a correlated equilibrium\footnote{This is actually a generalization of Defn.~\ref{def:correq} introducing a random variable $Z$ and changing the Markov chains to $X' - (X,Z) - Y$ and $Y' - (Y,Z) - X$.} that can be securely sampled by malicious parties (i.e., $I(X,Z; Y,Z) = C(X,Z; Y,Z)$).
Hence, the payoffs attainable by rational players with cheap talk are the same as those realizable via  secure sampling by malicious parties.
On the other hand, the payoffs in the biconvex-span of the Nash payoffs (attainable by rational players with polite talk) do not necessarily correspond to any Nash equilibria, and hence may not be realizable via secure sampling by malicious parties with polite talk.
However, all payoffs in the biconvex span of Nash payoffs are dominated by the Nash payoffs themselves, which are immediately attainable via a Nash equilibrium that can be securely sampled by malicious parties with polite talk.

\bibliographystyle{styles/IEEEtran}
\bibliography{references}

\end{document}